# FACTORS THAT INFLUENCE THE ADOPTION OF HUMAN-AI COLLABORATION IN CLINICAL DECISION-MAKING

*Research in Progress*

Patrick Hemmer[*], Karlsruhe Institute of Technology, Germany, patrick.hemmer@kit.edu

Max Schemmer[*], Karlsruhe Institute of Technology, Germany, max.schemmer@kit.edu

Lara Riefle, Karlsruhe Institute of Technology, Germany, lara.riefle@kit.edu

Nico Rosellen, IBM, Germany, nico.rosellen@ibm.com

Michel Vössing, Karlsruhe Institute of Technology, Germany, michael.voessing@kit.edu

Niklas Kühl, Karlsruhe Institute of Technology, Germany, niklas.kuehl@kit.edu

## Abstract

*Recent developments in Artificial Intelligence (AI) have fueled the emergence of human-AI collaboration, a setting where AI is a coequal partner. Especially in clinical decision-making, it has the potential to improve treatment quality by assisting overworked medical professionals. Even though research has started to investigate the utilization of AI for clinical decision-making, its potential benefits do not imply its adoption by medical professionals. While several studies have started to analyze adoption criteria from a technical perspective, research providing a human-centered perspective with a focus on AI's potential for becoming a coequal team member in the decision-making process remains limited. Therefore, in this work, we identify factors for the adoption of human-AI collaboration by conducting a series of semi-structured interviews with experts in the healthcare domain. We identify six relevant adoption factors and highlight existing tensions between them and effective human-AI collaboration.*

## 1 Introduction

Over the last few years, Artificial Intelligence (AI) has demonstrated capabilities similar to those of domain experts in many clinical decision-making tasks, e.g., the diagnosis of cancer in histopathology (Coudray et al., 2018), detecting diabetic retinopathy (Gargeya and Leng, 2017), or assessing X-ray scans for clinical findings such as pneumonia (Stephen et al., 2019). In line with these domain-specific advances are the evolving collaborative capabilities of AI, which open up the possibility for humans and AI to work together as coequal partners to improve medical diagnoses further (Dellermann et al., 2019; Lai et al., 2021; Seeber et al., 2020). From these developments emerged the notion that both have the potential to complement each other through different capabilities (Bansal et al., 2021; Zhou et al., 2021). For example, state-of-the-art AI can assist domain experts by efficiently analyzing large amounts of high-dimensional data—tasks that humans cannot perform due to the limits of human cognition (Dellermann et al., 2019). In contrast to when AI was mainly automating routine human tasks or used as a tool in the past, human-AI collaboration implies that AI works jointly with humans to solve problems (Seeber et al., 2020). We, therefore, define human-AI collaboration *"as an evolving, interactive process whereby two or more parties actively and reciprocally engage in joint activities aimed at achieving one or more shared goals"* (Lai et al., 2021, p. 390). In the context of clinical decision-making, this refers to medical personnel working together with AI—namely machine learning-based systems (Berente et al., 2021)—by bilaterally sharing information and collaboratively forming decisions based on interactive knowledge exchange. By now, an increasing number of studies have shown the potential of human-AI collaboration in the medical domain (Cai et al., 2019; Esteva et al., 2021; Hekler et al., 2019; Lai et al., 2021; Tschandl et al., 2019). For example, Tschandl et al. (2019) demonstrate superior skin cancer diagnosis accuracy of physicians collaborating with AI over that of either AI or physicians alone.

---

[*] Both authors contributed equally in a shared first authorship.

Even though general adoption factors of information systems have a long research history (Davis, 1993; Venkatesh et al., 2003, 2016) and several studies have investigated the adoption of AI as a tool in healthcare, e.g., wearables (Park, 2020) or chatbots (Laumer et al., 2019), the adoption of human-AI collaboration in clinical decision-making still remains scarce in practice mainly due to missing user-centeredness during its design process as well as the opaqueness of AI algorithms (Beede et al., 2020; Cabitza et al., 2017; Cai et al., 2019; Khairat et al., 2018; Lee et al., 2021; Wang et al., 2021). Contrary to many other domains, decision-making in the clinical context directly affects human lives and is accompanied by high costs in case of errors. Despite the potential of human-AI collaboration, physicians might, in the end, be still held accountable for their final decision-making regardless of being advised by AI (Durán and Jongsma, 2021). Moreover, it takes years of training and experience for humans to acquire the necessary knowledge in the medical domain, as making a diagnosis is often a highly complex process that involves considering various disease patterns (Bunniss and Kelly, 2010). For these reasons, it cannot be taken for granted that experts recognize the utility of AI as a trustworthy teammate and are open to collaboration (Seeber et al., 2020). As a current structured literature review highlights the need for research regarding the adoption of human-AI collaboration in a clinical setting (Lai et al., 2021), we derive the following research question in this paper:

**RQ:** *What are relevant factors that influence the adoption of human-AI collaboration in clinical decision-making?*

Since human-AI collaboration is an emerging concept, we approach the research question from a qualitative and exploratory perspective and conduct a series of semi-structured interviews to collect insights into the adoption of human-AI collaboration in clinical decision-making. The ten interviewees are experts from the field of radiology, pneumology, pathology, and AI. Through inductive coding, we derive six adoption factors. Professionals state the need for a *complementary* AI that communicates its insights *transparently* and *adapts* to the users to enable *mutual learning* and *time-efficient* work with a final *human agency*. When comparing these adoption factors with existing literature on effective human-AI collaboration, several areas of tension are found. For example, our study shows that medical professionals desire human agency, which does not necessarily always result in the best possible human-AI collaboration performance (Kerrigan et al., 2021; Wilder et al., 2021). Therefore, future research needs to find ways to create effective but also accepted forms of human-AI collaboration. With this paper, we contribute to the body of knowledge by identifying relevant factors for the adoption of human-AI collaboration in clinical decision-making.

The remainder of this paper is structured as follows. In Section 2, we elaborate on the foundations of this work and refer to related literature. In Section 3, we outline our research methodology. Subsequently, we present the identified factors for the adoption of human-AI collaboration in clinical decision-making in Section 4. Next, we discuss these factors by contrasting them to the existing literature on effective human-AI collaboration in Section 5. Lastly, we summarize our contributions and present an outlook of the planned next steps of our research in Section 6.

## 2 Foundations and Related Work

Clinical decision support systems (CDSS) are computer systems that have been designed to improve patient care. The first CDSS have already been introduced over four decades ago (Kaplan, 2001).

Initially, these systems were an advancement of early expert systems to simulate human thinking (Berner, 2007; Gustafson et al., 1992; Pingree et al., 1993). CDSS have been developed to support decision-making before, during, and after clinical procedures in a variety of contexts, e.g., highlighting potential adverse drug reactions and allergies (Moxey et al., 2010), drug prescription checking (Berner, 2007), or proposing diagnostic hypotheses based on clinical manifestations (Barnett et al., 1987).

In recent years, with the constantly growing capabilities of AI, this technology has soon been considered for the development of CDSS (Patel et al., 2009) as it has demonstrated remarkable results in, e.g., computer vision (He et al., 2015) and natural language processing tasks (McClelland et al., 2020). When combined with large data sets, AI is expected to outpace the use of traditional CDSS soon (Middleton et al., 2016). Additionally, from the constantly growing capabilities of AI evolved the notion that CDSS will soon form a collaboration to work closely together with clinical experts and perform tasks jointly (Grudin, 2017; Seeber et al., 2020), which we refer to as AI-based CDSS. This idea stems from the research direction of human-AI collaboration (Seeber et al., 2020; Zschech et al., 2021). It aims to establish a socio-technical ensemble to extend each other's capability limits by leveraging the respective strengths of humans and AI, resulting in superior task outcomes (Bansal et al., 2021; Dellermann et al., 2019). For example, radiologists' capabilities can be augmented through AI-based CDSS that provide them with insights acquired through the analysis of large amounts of high-dimensional data, e.g., MRI, CT, or X-ray scans, or through proposing diseases from historical cases about which inexperienced physicians might not be aware (Patel et al., 2009; Zhou et al., 2021). Vice versa, physicians' knowledge can be incorporated into the AI-based CDSS to contribute to steady model improvements (Hemmer et al., 2020; Mahapatra et al., 2018).

However, the adoption of human-AI collaboration in clinical decision-making remains limited in clinical practice (Lai et al., 2021). Possible reasons for the low adoption of these systems are missing user-centered design relating to the relevance of the information provided by the system, its perceived validity as well as the opaqueness of many AI algorithms (Baig et al., 2017; Khairat et al., 2018; Lee et al., 2021). These factors strongly relate to adoption theories such as the technology acceptance model (TAM) by Davis (1993), which has been refined by Venkatesh et al. (2003) in the unified theory of acceptance and use of technology (UTAUT). They aspire to provide potential explanations for how anticipation about performance, social influences, and facilitating conditions are determinants of a system's acceptance and usage (Venkatesh et al., 2003). In this context, several studies have already leveraged these frameworks for studying adoption factors. For example, Van Schaik et al. (2004) study the acceptance of gastroenterology referral CDSS and find that performance expectancy is the most meaningful predictor of technology acceptance. Other studies report reasons for low acceptance of evaluated CDSS due to interferences with the workflow (Curry and Reed, 2011), presentation of already known information (Terraz et al., 2005), or low confidence in generated evidence (Sousa et al., 2015).

Nevertheless, knowledge regarding the adoption of human-AI collaboration has largely been unexplored so far as previous work a) has predominantly investigated non-AI-based CDSS without considering them as equal collaboration partners and b) has focused its evaluation mainly on technical facets while neglecting user-centeredness concerning its adoption in clinical practice (Lai et al., 2021). Therefore, we aim to address this research gap in this work by deriving adoption factors from a user-centered point of view.

## 3 Methodology

To identify factors that influence the adoption of human-AI collaboration in healthcare, we pursued an inductive approach collecting qualitative data (Eisenhardt and Graebner, 2007). We conducted semi-structured in-depth interviews with ten purposefully sampled physicians and experts on AI in healthcare (Coyne, 1997). The interviewees were chosen based on their active involvement in clinical decision-making—in the case of the five physicians including assistant, senior, and chief physicians—or their diverse expertise on user requirements for AI-based CDSS—in the case of the five AI experts. The experts on AI in healthcare have been implementing AI-based CDSS in collaboration with many physicians and, thus, acquired a broad knowledge of their specific needs.

Interviews lasted between 25 to 50 minutes and were conducted and recorded using Webex. Before starting with the interview questions, a common understanding of human-AI collaboration in clinical decision-making as previously defined was ensured. As the first part of the interview, interviewees were asked about their specific role in the context of clinical decision-making to tailor subsequent questions to their role and specific area of their involvement.

Generally, the overarching topics were identical for all interviewees. The interview questions covered the process of decision-making in the clinical context, particularly critical decision stages, currently used support systems, as well as possible use cases for collaboration with AI. Particular emphasis was put on the factors that would need to be given for the medical professionals to collaborate with the AI. By focussing on open-ended questions, we ensured that interviewees could express themselves freely (Gläser and Laudel, 2009).

Interview transcripts consist of 89 pages and are the foundation for the following analysis. We applied qualitative content analysis according to the recommendations of Mayring (2015) to analyze the gathered data. In detail, we pursued an inductive approach based on an iterative process of paraphrasing, reducing, and aggregating statements relevant to our research question using MAXQDA. In particular, we first extracted interview statements that provided insights on adoption factors for human-AI collaboration and paraphrased them.

We found 327 statements relevant to our research question that were then analyzed by coding them according to the two-cycle recommendations by Saldaña (2021). In the first cycle, the main topic of each statement (e.g., avoiding disturbing interruptions of clinical staff) was summarized using descriptive coding. For example, when an interviewee stated that a CDSS would need to give "*physicians more time to really take care of the critical things and the patient*" (I6), the statement was coded as "*need for improvement of time efficiency*". Next, pattern coding was applied in the second coding cycle to aggregate the descriptive first-cycle codes to a higher level of abstraction to derive major themes in the data. For example, codes reflecting the need for time efficiency were aggregated together with codes reflecting the wish for easy-to-use systems.

All codes were discussed by three researchers until consensus was reached. Thereby, six factors influencing the adoption of human-AI collaboration in healthcare were identified.

## 4 Factors for the Adoption of Human-AI Collaboration in Clinical Decision-Making

To facilitate the adoption of human-AI collaboration in clinical practice, several key factors have to be taken into consideration to exploit AI's existing potential for successful collaboration with medical professionals. In this section, we present the identified factors relevant for the adoption of human-AI collaboration in clinical decision-making.

A central factor through which AI can add significant value in collaboration with medical professionals is to assist them through **complementary** capabilities. Medical professionals see the most added value in the potential that AI can complement them by "*improving and facilitating their stressful work*" (I10) in those tasks that are especially difficult for them in particular or for humans in general. Hereby, AI can not only relieve medical experts from repetitive tasks but also offer assistance when it comes to highly complex cases. In this context, it can provide, e.g., a second medical opinion for ambiguous clinical findings by recognizing patterns in high-dimensional data or supporting repetitive work that still requires a high degree of cognitive focus. For instance, one interviewee from the field of pathology reported that "*after 20 or 30 examinations during a day, eyes become tired resulting in less concentration in the evening*" (I5). Moreover, it could support physicians in differential diagnosis by weighting the probability of one disease versus that of other diseases when accounting for a patient's illness or "*by proposing likely diseases from historical cases about which medical experts were not aware of*" (I2). In contrast, from a team point of view, physicians can complement AI through interpersonal tasks, "*show empathy, consider boundary conditions and external factors, and classify these in a holistic cross-domain picture*" (I10).

Another emerging factor is **mutual learning**. Interviewees emphasize the possibility for mutual improvement through bilateral feedback loops. On the one hand, medical experts can add "*new insights and training data to update AI models*" (I9) to increase their accuracy over time. Likewise, AI-initiated "*feedback loops can enable physicians to improve*" as, for instance, radiologists "*rarely receive feedback on their decisions*" (I2) since they are mostly responsible for the patient at the moment of diagnosis and not beyond. If further examinations reveal a different finding than initially diagnosed, a digital health record and AI-initiated feedback could also contribute to learning from the initially mistaken expert. Currently, "*a lot of learning potential gets lost, which could improve diagnostic accuracy*" (I2) even further.

**User adaptiveness** marks another important factor that could be identified during the interviews. As in practice, multiple user groups are likely to collaborate with an AI-based CDSS, the system has to account for their respective characteristics and adapt accordingly. This means that it has to be clear "*who this solution is actually for, and then these people have to be brought on board*" (I7). For instance, young assistant physicians who have less professional experience require different levels of support. Similar is the case with nurses and medical technical assistants. "*In case no senior physician is available, which might happen during night shifts, assistant physicians have to decide on their own*" (I6). Hence, AI-based CDSS should account for differing needs by varying levels of information presentation dependent on the knowledge and background of their collaborators.

Our study highlights the necessity for medical professionals to understand the AI's decision-making as it has been stated that "*it is important that the AI can explain its recommendations*" (I3). The explanations need to be "*presented in a comprehensible way without a high level of technical understanding*" (I6) to enable efficient usage during clinical practice, which can be considered as **decision transparency**. Additionally, medical professionals also highlighted that explanations should be adaptive. It could be that they initially want to "*have a detailed explanation*" (I3) but later "*only want detailed explanations the first and second time*" (I3) and no explanations after they have "*learned that they can trust the recommendations*" (I3).

Furthermore, the interviews revealed that **time efficiency** is a critical adoption factor for medical professionals. They mentioned that a lack of time is one of the most important stressors in their daily work. Therefore, time that is consumed by collaboration with AI needs to be kept to a minimum. AI should not disturb medical professionals in their daily work, as "*there is simply not enough time for this, and acceptance would decrease*" (I7). Moreover, it is crucial that the collaboration is integrated into existing workflows and that it does not "*bother and distract*" (I7) physicians. Conversely, the system should be integrated into the workflow "*with simultaneous user-friendliness or simplicity of operation*" (I8) to ensure efficient work. Furthermore, medical professionals highlighted that human-AI collaboration should not only be seamless and efficient itself but should overall also save time, e.g., by filtering out easy cases as one doctor stated, "*of course, it would be very helpful if the machine no longer displayed inconspicuous images so that you would only get relevant images that are important for the diagnosis*" (I3). They even argued that the AI should even take over administrative tasks in the best case. This could have the effect "*that physicians would be given more time to really take care of the critical things and the patient*" (I6).

Lastly, it is essential that clinical personnel are actively involved in the decision-making process and can "*intervene at any time in case the physician's opinion is contradicting to the recommendation of the system*" (I1) so that the final decision and **agency** remains with the human expert. Even if tasks are not critical and monotonous, medical experts still highlighted the need for a final human quality control: "*I think especially for the less demanding routine work, a lot of it can be done by AI, with a final quality control that you just have to click again*" (I8). This is related to the debate that physicians will be held accountable also for automated decisions that would have been conducted under their supervision (Durán and Jongsma, 2021).

Table 1 displays exemplary quotes from the expert interviews. Interestingly, when analyzing the interview results, no area of tension arises between both expert groups as all resulting factors could be identified in both groups.

| Factors | Exemplary quotes from the expert interviews |
|---|---|
| **Complementarity** | *"The support provided by AI is certainly beneficial and can improve and facilitate the stressful work of physicians. However, especially the holistic picture and external influencing factors and boundary conditions, those are actually the points that physicians can add to the decision-making that the machine is not able to incorporate"* (I10). |
| **Mutual Learning** | *"If we have an AI-based feedback loop, it would be very plausible for me that physicians improve because they rarely receive feedback on their decisions. Sometimes you learn that something different came out than you would have initially thought. However, you often do not see the patients again, and then you just do not notice it. That is a pity, as a lot of learning potential gets lost, which could improve diagnostic accuracy"* (I2). |
| **User Adaptiveness** | *"Not only physicians can be supported with AI, but also nurses or medical technical assistants. For example, nurses have to decide if the heart rhythm is all right and if steps should be taken and medication given. Or in case no senior physician is available, which might happen during night shifts, assistant physicians have to decide on their own"* (I6). |
| **Decision Transparency** | *"I think it is important that the AI can explain its recommendations. This will probably not be necessary for cases that are clear. However, in edge cases where there is a discrepancy, it is important to understand the criteria that the decision is based on"* (I3). |
| **Time Efficiency** | *"When the doctor has little time, if you bother and distract the doctor with something like that all the time, then I think you lose acceptance very quickly. Especially, when it comes to things that are not acutely life-threatening"* (I2). |
| **Agency** | *"I think that it is certainly important that one can also intervene in the process, so that the final diagnosis is conducted by medical professionals who are responsible for it"* (I5). |

*Table 1.     Identified factors for the adoption of human-AI collaboration in clinical decision-making including exemplary quotes from the interviews.*

## 5   Discussion

This section discusses the factors identified from the interviews by relating and contrasting them to the existing literature on effective human-AI collaboration. Regarding the identified factors in this study, we find desiderata that align with effective human-AI collaboration and desiderata that create areas of tension.

In line with the adoption factor of **complementarity**, Dellermann et al. (2019) argue that humans and AI generally possess different potentially complementary skills. However, its realization in terms of superior collaborative task performance often remains an unsolved problem in practice (Hemmer et al., 2021). A possible reason might be that humans do not always appropriately rely on AI advice and might even be convinced of explanations in cases of incorrect recommendations (Bansal et al., 2021). Nevertheless, it became apparent in the discussions with physicians that they are open to AI support. In particular, it can help them illuminate complex decisions through a different complementary perspective or provide a second opinion on ambiguous cases, positively affecting the adoption of human-AI collaboration. Thus, we can conclude that the adoption factor of the medical professionals aligns with a central requirement for effective human-AI collaboration.

**Mutual learning** is a facet with increasing benefits over time, as mentioned by several studies (Gu et al., 2021; Lee et al., 2021). If this potential becomes obvious to medical professionals, e.g., in the form of an apparent knowledge gain initiated by relevant information from the AI, this factor can contribute to fostering the adoption of AI in the long term. Moreover, human feedback can be used to improve the AI's decision quality over time by providing it with verified instances that contribute to its learning process the most (Hemmer et al., 2020). In this context, Gu et al. (2021) integrate an AI learning mechanism in a diagnosis tool for pathologists, which provides motivating feedback on how experts' knowledge could improve the AI. Mutual learning also includes the AI teaching medical professionals. While this might be possible to a certain degree, e.g., by highlighting relevant information physicians might not have been aware of, complex decision rules derived by AI might be impossible to teach to humans (Adadi and Berrada, 2018). This could eventually lead to an area of tension as the medical professional desires to learn something from the AI, e.g., how the AI arrived at its decision but is not able to understand it.

Research on human-AI collaboration has highlighted the benefits and needs of **user adaptiveness** of AI with regard to adapting to humans' strengths and weaknesses (Keswani et al., 2021; Wilder et al., 2021) or adapting its explanations with regard to information presentation (Bansal et al., 2021; Kühl et al., 2020). In this context, Nourani et al. (2020) find that users' domain knowledge affect their trust in the AI system. Moreover, adapting AI not only to users' knowledge but also to their individual characteristics, e.g., cognitive styles, is likely to foster human acceptance (Riefle and Benz, 2021; Riefle et al., 2022). Therefore, adapting AI's behavior to different knowledge levels and other factors might benefit human-AI collaboration. We can conclude that the adoption factor of the medical professionals aligns with existing literature on effective human-AI collaboration.

Our interview study has additionally highlighted the need for AI's **decision transparency**. Current research shows that transparency needs to be carefully designed to result in effective collaboration (Bansal et al., 2021). If designed appropriately, transparency can lead to a calibrated level of trust, resulting in improved collaboration performance (Zhang et al., 2020). Therefore, we conclude that the desideratum of the medical professionals aligns with the identified criteria for effective human-AI collaboration.

**Time efficiency** is one of the most often highlighted adoption factors of medical professionals. This is an interesting factor as human-AI collaboration does not necessarily improve time efficiency. Collaboration, per definition, still requires human input. Some forms of human-AI collaboration may even increase the time that is needed to derive a final decision. For example, if the medical professionals change from intuitive decision-making to analytical processing (Buçinca et al., 2021; Jussupow et al., 2021), the time needed to process the information might increase. Current research sees the benefits in human-AI collaboration mostly in increased task effectiveness (Bansal et al., 2021), which does not necessarily align with the desideratum of the professionals and is an area of tension that needs to be addressed in future research.

Finally, our interview study highlights that medical professionals demand the final **agency** in the decision-making process even if the tasks are not critical. In general, this might be driven by the fact that they are held accountable for their decision-making (Durán and Jongsma, 2021). However, research has demonstrated that in certain cases leaving the final decision to AI can improve the overall process effectiveness (Keswani et al., 2021; Wilder et al., 2021). For example, research on forecasting theory has shown that taking averages of the decision of model and human can result in superior performance compared to both making the decision on their own (Bunn and Wright, 1991; Blattberg and Hoch, 2010). In this case, an algorithmic aggregation mechanism derives the final decision. Other researchers have shown the benefits of a priori distributing tasks (Keswani et al., 2021; Wilder et al., 2021). However, the medical professionals would have to give up some of their decision-making power in this setting. Therefore, we see an area of tension that should be analyzed in future research.

## 6 Conclusion and Future Work

This study identifies factors influencing the adoption of human-AI collaboration in clinical decision-making. Taking a user-centered perspective, we conducted a series of ten semi-structured interviews with domain experts. The analysis revealed six adoption factors for human-AI collaboration in clinical decision-making. Professionals state the need for a *complementary* AI that communicates its insights *transparently* and *adapts* to the users to enable *mutual learning* and *time-efficient* work with a final *human agency*. Subsequently, we contrasted them with existing literature on effective human-AI collaboration. While some factors fit into the prevailing concepts, our work has identified areas of tension in others that can provide an explanation for why adoption challenges may arise after the deployment of AI-based CDSS in practice.

Despite providing valuable insights, this study is not without limitations. Currently, our study is limited to ten interviews. However, its interviewees encompass senior and chief medical professionals from multiple hospitals and sectors. Furthermore, due to its focus on the healthcare domain, future research is needed to evaluate the generalizability of the identified adoption factors in other high stake decision-making areas (e.g., the legal domain).

Some of our identified factors can be mapped to existing work and theories on adoption. For example, the complementarity of humans and AI can be interpreted as an expected performance improvement which is one of the key constructs in TAM and UTAUT (Im et al., 2011). However, other factors may be an extension as well. Therefore, we plan to synthesize literature and our identified user-centered factors in future work. Following that, we want to validate the extended set of adoption factors by conducting focus groups with medical professionals and AI experts. Finally, we plan to conduct a quantitative evaluation with our refined and validated set of adoption factors. More specifically, we aim to analyze the influence of the identified factors on the effectiveness of human-AI collaboration through a user survey quantitatively. Our findings can provide a starting point for deriving guidance for the successful design and deployment of AI-based CDSS. Building on the factors presented in this study, researchers and practitioners may derive design guidelines for future AI-based CDSS to foster accepted and successful human-AI collaboration.